\documentclass{article}

\usepackage[english]{babel}
\usepackage[latin1]{inputenc}
\usepackage{amsmath, amstext, amssymb, amsthm, url, hyperref, pst-all}

\newcommand{\mc}[1]{\mathcal{#1}}
\newcommand{\mbb}[1]{\mathbb{#1}}
\newcommand{\msf}[1]{\mathsf{#1}}
\newcommand{\card}[1]{\overline{\overline{{#1}}}}
\newcommand{\urlbib}[1]{{\scriptsize{\url{http://#1}}}}

\newcommand{\ket}[1]{\ensuremath{|\,#1\,\rangle}}

\newcommand{\fp}[1]{#1 \negthickspace :}
\newcommand{\eqdef}{\stackrel{\text{\tiny \upshape def}}{=}}

\newtheorem{definition}{Definition}
\newtheorem{example}{Example}
\newtheorem{theorem}{Theorem}
\newtheorem{corollary}{Corollary}

\title{Paraconsistent Machines and their Relation to  Quantum Computing}
\author{Juan C. Agudelo\footnote{Ph.D. Program in Philosophy, area of Logic, IFCH and Group for Applied and Theoretical Logic- CLE, State University of Campinas - UNICAMP, Brazil. Logic and Computation Research Group, Eafit University, Colombia.} \and Walter Carnielli\footnote{IFCH and  Group for Applied and Theoretical Logic- CLE, State University of Campinas - UNICAMP, Brazil. Security and Quantum Information Group- IT, Lisbon, Portugal.}}
\date{}

\begin{document}

\maketitle

\begin{abstract}
We describe a method  to  axiomatize  computations in  deterministic
Turing machines. When applied to computations in
non-deterministic  Turing machines,  this method may produce
contradictory  (and therefore trivial) theories,  considering
classical logic as the underlying logic.  By substituting in such
theories the underlying logic by  a paraconsistent logic
we define  a new computation model,  the
 \emph{paraconsistent Turing machine}. This
model allows a partial simulation of superposed states of
quantum computing.  Such a feature  allows the  definition  of
paraconsistent algorithms  which solve (with some restrictions)
the well-known Deutsch's and Deutsch-Jozsa
problems.  This first  model of computation, however,
does not adequately represent the notions of \emph{entangled
states} and \emph{relative phase}, which are key features in quantum computing.
In this way,  a more sharpened model of paraconsistent Turing machines  is defined,
which better approaches quantum computing features.
Finally, we define complexity classes for such models,  and establish some
relationships  with classical complexity classes.
 \end{abstract}

\section{Introduction}

The undecidability of first-order logic was first proved by
Alonzo Church in \cite{Church-1936b} and an alternative proof with
the same  result was presented by Alan Turing in
\cite{Turing-1936}. In his paper, Turing defined an abstract
model of automatic machines, now known as \emph{Turing machines}
(TMs), and demonstrated that there are unsolvable problems for that
class of machines. By axiomatizing machine
computations in first-order theories, he could then prove  that
the decidability of first-order logic  would  imply the solution
of established unsolvable problems. Consequently, by
\emph{reductio ad absurdum}, first-order logic is shown to be
undecidable. Turing's proof was simplified by Richard B\"uchi in
\cite{Buchi-1962}, and a more recent and clear version of this
proof is presented by George Boolos and Richard Jeffrey in
\cite[Chap. 10]{Boolos-Jeffrey-1989}.

By following \cite{Boolos-Jeffrey-1989} and adding new axioms, we
define a method to obtain adequate theories for computations in
\emph{deterministic} TMs  (DTMs)  (Sec. \ref{axio-TM-comp}), which
verify a formal notion of representation of TM computation (here introduced),
therefore enhancing the standard way in which TMs are expressed by means of
classical first-order logic.

Next, we will show that by using the same axiomatization method for
\emph{non-deterministic} TMs  (NDTMs), we obtain (in some cases)
contradictory theories, and therefore trivial theories in view  of the
underlying logic. At this point, we have two options in sight to
avoid triviality: (a) The first option, consisting in the classical
move of  restricting theories  in a way that contradictions could
not be derived (just representing computations in NDTMs); (b) the
second option, consisting in substituting the underlying logic by
a paraconsistent logic, supporting contradictions and providing a way to define new models of
computation through the interpretation of the theories. The first
option is sterile: incapable of producing a new model of
computation. In this paper, we follow the second option,
initially defining in Sec. \ref{ptms} a model of
\emph{paraconsistent} TMs (ParTMs), using the paraconsistent
logic $LFI1^*$ (see \cite{Carnielli-Coniglio-Marcos-2007}). We
then show that ParTMs allow a partial simulation of `superposed states', an important
feature of quantum computing (Sec. \ref{sim-qc-ptms}). By using this property,
and taking advantage of `conditions of inconsistency' added to the instructions,
we show that the quantum solution of Deutsch's and
Deutsch-Jozsa problems can be simulated in ParTMs (Sec.
\ref{sim-D-DJ-prob}). However, as explained in Sec.
\ref{sim-ent-states-rel-phases}, ParTMs are not adequate to simulate
\emph{entangled states} and \emph{relative phases}, which are key features in quantum computing.
Thus, still in Sec. \ref{sim-ent-states-rel-phases}, we define a
paraconsistent logic with a \emph{non-separable} conjunction and,
with this logic, a new model of paraconsistent TMs is defined,
which we call \emph{entangled paraconsistent} TMs (EParTMs).
In EParTMs, uniform entangled states can be successfully
simulated. Moreover, we describe how the notion of relative phase (see \cite[p.
193]{Nielsen-Chuang-2000}) can be introduced
in this model of computation.\footnote{ParTMs were first presented in \cite{Agudelo-Sicard-2004}
and relations with quantum computing were presented in \cite{Agudelo-Carnielli-2005},
but here we obtain some improvements and introduce the model of EParTMs,
which represents a better approach to quantum computing.}

The \emph{paraconsistent computability theory} has already been
mentioned in \cite[p. 196]{Sylvan-Copeland-2000} as an `emerging
field of research'. In that paper, \emph{dialethic machines} are
summarily described as Turing machines acting under dialethic
logic (a kind of paraconsistent logic) in the presence of a
contradiction, but no definition of any computation model is
presented.\footnote{The authors say  that ``it is not difficult to
describe how a machine might encounter a contradiction: For some
statement $A$, both $A$ and $\neg A$ appear in its output or
among its inputs'' (cf. \cite[p. 196]{Sylvan-Copeland-2000});
but how can $\neg A$ `appear'? They also claim that ``By contrast
[with a classical machine], a machine programmed with a dialethic
logic can proceed with its computation satisfactorily [when a
contradiction appears]''; but how would  they proceed?} Also,
an obscure argument is presented in an attempt to show how dialethic machines
could be used to compute classically uncomputable functions.
Contrarily, our models of ParTMs and EParTMs do not intend to
break the Church-Turing thesis, i.e., all problems computed by
ParTMs and EParTMs can also be computed by TMs (Sec.
\ref{comp-power-partms-EParTMs}). However, such models  are
advantageous for the understanding of quantum computing and
parallel computation in general. Definitions of computational
complexity classes for ParTMs and EParTMs, in addition to interesting
relations with existing classical and quantum computational
complexity classes, are presented in Sec. \ref{comp-power-partms-EParTMs}.

The paraconsistent approach to quantum computing presented here
is just one way to describe  the role of quantum features in the
process of computation by means of non-classical logics; in
\cite{Agudelo-Carnielli-2007} we presented another way to define
a model of computation based on another paraconsistent logic, also
related  with quantum computing. The relationship  between these
two different definitions of paraconsistent computation is a task
that needs to be addressed in future work.

\section{Axiomatization of TM Computations}\label{axio-TM-comp}

As mentioned above, a method to define first-order theories
for Turing machine computations has already been introduced in \cite{Turing-1936}.
Although this is a well-known construction,  in view of the important role of
this method in our definition of paraconsistent Turing machines  we will
describe it in detail, following \cite[Chap. 10]{Boolos-Jeffrey-1989}.
This method will be extended to deal with a formal notion of representation (here introduced) of TM computation.

Considering $Q = \{q_1, \ldots, q_n \}$ as a finite set of states
and $\Sigma = \{s_1, \ldots, s_m \}$ as a finite set of
read/write symbols, we will suppose that TM instructions are defined by
quadruples of one of the following types (with the usual interpretations, where
$R$ means a movement to the right,  and $L$ means a movement to the left):
\begin{align}
    &q_i s_j s_k q_l,\tag{I}\label{inst-i}\\
    &q_i s_j R q_l,\tag{II}\label{inst-ii}\\
    &q_i s_j L q_l.\tag{III}\label{inst-iii}
\end{align}

By convention, we will enumerate the instants of time and
the cells of the tape by integer numbers, and we will consider that
machine computations begin at time $0$, with a symbol sequence on
the tape (the \emph{input} of the computation), and with the
machine in state $q_1$ scanning the symbol on cell $0$. $s_1$ will be
assumed to be an \emph{empty} symbol.

In order to represent the computation of a TM $\mc{M}$ with  input
$\alpha$ (hereafter $\mc{M}(\alpha)$), we initially define the
first-order theory $\Delta_{FOL}(\mc{M}(\alpha))$ over the first-order language $\mc{L}
= \{Q_1, \ldots, Q_n, S_1, \ldots, S_{m}, <, ', 0\}$,\footnote{The
subscript $FOL$ on $\Delta$ aims to  emphasize the fact that we
are considering the \emph{classical first-order logic} (FOL) as
the underlying logic of the theory, i.e. $\Delta_{FOL} \vdash A$
means $\Delta \vdash_{FOL} A$. A different subscript will
indicate that another (non-classical) first-order logic is being
taken into consideration.} where
symbols $Q_i$, $S_j$ and $<$ are binary predicate symbols, $'$ is
a unary function symbol and $0$ is a constant symbol. In the
intended interpretation $\mc{I}$ of the sentences in
$\Delta_{FOL}(\mc{M}(\alpha))$, variables are interpreted as
integer numbers, and symbols in $\mc{L}$ are interpreted in the
following way:
\begin{itemize}
    \item $Q_i(t, x)$ indicates that $\mc{M}(\alpha)$ is in state $q_i$, at time $t$,
    scanning the cell $x$;
    \item $S_j(t, x)$ indicates that  $\mc{M}(\alpha)$ contains  the  symbol $s_j$, at
     time $t$, on cell $x$;
    \item $<(x, y)$ indicates that $x$ is less than $y$, in the standard order of
     integer numbers;
    \item $'(x)$ indicates the successor of $x$;
    \item $0$ indicates the number $0$.
\end{itemize}
To simplify notation, we will use $x < y$ instead of $<(x, y)$ and
$x'$ instead of $'(x)$. The theory $\Delta_{FOL}(\mc{M}(\alpha))$
consists of the following axioms:
\begin{itemize}
    \item Axioms  establishing  the properties of $'$ and $<$:
        \begin{align}
            &\forall z \exists x (z = x'), \tag{A1} \label{existe-suc}\\
            &\forall z \forall x \forall y (((z=x') \wedge (z=y')) \to (x=y)), \tag{A2} \label{unicidad-suc}\\
            &\forall x \forall y \forall z (((x<y) \wedge (y<z)) \to (x<z)), \tag{A3} \label{trans-menorque}\\
            &\forall x (x < x'), \tag{A4} \label{relac-suc-menorque}\\
            &\forall x \forall y ((x<y) \to (x \neq y)). \tag{A5} \label{antireflex-menorque}
        \end{align}
    \item An axiom for each  instruction $i_j$ of $\mc{M}$. The axiom is defined depending
    respectively on the instruction type  \eqref{inst-i}, \eqref{inst-ii} or \eqref{inst-iii}
     as:
            \begin{multline}
                    \forall t \forall x \Biggl(\biggl(Q_i(t, x) \wedge S_j(t, x)\biggr) \to \biggl(Q_l(t', x) \wedge S_k(t', x) \wedge \\ \forall y \Bigl((y \neq x) \to
                    \Bigl(\bigwedge_{i=1}^{m}\bigr(S_i(t, y) \to S_i(t', y)\bigr)\Bigr)\Bigr)\biggr)\Biggr), \tag{A$i_{\msf{j}}$ \eqref{inst-i}} \label{ax-inst-i}
                \end{multline}
                \begin{equation}
                    \forall t \forall x \Biggl(\biggl(Q_i(t, x) \wedge S_j(t, x)\biggr) \to \biggl(Q_l(t', x') \wedge \\ \forall y \Bigl(\bigwedge_{i=1}^{m}\bigl(S_i(t, y) \to
                    S_i(t', y)\bigr)\Bigr)\biggr)\Biggr), \tag{A$i_{\msf{j}}$ \eqref{inst-ii}} \label{ax-inst-ii}
                \end{equation}
                \begin{equation}
                    \forall t \forall x \Biggl(\biggl(Q_i(t, x') \wedge S_j(t, x')\biggr) \to \biggl(Q_l(t', x) \wedge \\ \forall y \Bigl(\bigwedge_{i=1}^{m}\bigl(S_i(t, y) \to
                    S_i(t', y)\bigr)\Bigr)\biggr)\Biggr). \tag{A$i_{\msf{j}}$ \eqref{inst-iii}} \label{ax-inst-iii}
                \end{equation}
    \item An axiom to specify  the initial configuration of the machine. Considering the input
     $\alpha = s_{i_0} s_{i_1} \ldots s_{i_{p-1}}$, where $p$ represents the length of
      $\alpha$, this axiom is defined by:
        \begin{multline}
            Q_1(0, 0) \wedge \left(\bigwedge_{j=0}^{p-1} S_{i_{j}}(0, 0^{j})\right) \wedge \forall y \left(\left(\bigwedge_{j=0}^{p-1} y \neq 0^{j}\right) \to S_1(0, y)\right), \tag{A$\alpha$} \label{init-conf}
        \end{multline}
        where $0^{j}$ means  $j$  iterations of the successor ($'$) function to constant $0$.
\end{itemize}

In \cite{Boolos-Jeffrey-1989},  a sentence $H$ is defined to
represent the halting of the computation, and it is  thus proved
that $\Delta_{FOL}(\mc{M}(\alpha)) \vdash H$ iff the machine
$\mc{M}$ with input $\alpha$ halts. In this way, the decidability
of first-order logic implies the solution for the \emph{halting
problem}, a well-known unsolvable problem; this  proves  (by
\emph{reductio ad absurdum}) the undecidability of first-order
logic. For Boolos and Jeffrey's aims,
$\Delta_{FOL}(\mc{M}(\alpha))$ theories are strong enough, but
our purpose here is to attain  a precise logical representation of TM
computations. Therefore, we will formally  define the notion of
representability of a TM computation and show that new axioms must
be added  to $\Delta_{FOL}(\mc{M}(\alpha))$ theories. Our
definition of the representation of a TM computation (Definition
\ref{def-rep-comp}) is founded upon the definitions of the
representation of functions and relations (Definition
\ref{def-rep-func} and \ref{def-rep-rel}) in theories introduced
by Alfred Tarski in collaboration with Andrzej Mostowski and
Raphael M. Robinson in \cite{Tarski-Mostowski-Robinson-1953}.

\begin{definition}\label{def-rep-func}
    Let $f$ be a function of arity $k$, $\Delta$ an arbitrary theory and $\varphi(x_1, \ldots, x_k, x)$ a wff (with $k + 1$ free variables) in $\Delta$. The function $f$ is \emph{represented} by $\varphi$ in $\Delta$ if $f(m_1, \ldots, m_k) =  n$ implies (bars are used to denote numerals):
    \begin{enumerate}
        \item $\Delta \vdash \varphi(\bar{m_1}, \ldots, \bar{m_k}, \bar{n})$,\label{def-rep-func-cond-i}
        \item if $n \neq p$ then $\Delta \vdash \neg \varphi(\bar{m_1}, \ldots, \bar{m_k}, \bar{p})$, and\label{def-rep-func-cond-ii}
        \item $\Delta \vdash \varphi(\bar{m_1}, \ldots, \bar{m_k}, \bar{q}) \rightarrow \bar{q} = \bar{n}$.\label{def-rep-func-cond-iii}
    \end{enumerate}
\end{definition}

\begin{definition}\label{def-rep-rel}
    Let $R$ be a relation of arity $k$, $\Delta$ an arbitrary theory and $\varphi(x_1, \ldots, x_k)$ a wff (with $k$ free variables) in $\Delta$. The relation $R$ is \emph{represented} by $\varphi$ in $\Delta$ if:
    \begin{enumerate}
        \item $(m_1, \ldots, m_k) \in R$ implies $\Delta \vdash \varphi(\bar{m_1}, \ldots, \bar{m_k})$, and\label{def-rep-rel-cond-i}
        \item $(m_1, \ldots, m_k) \notin R$ implies $\Delta \vdash \neg \varphi(\bar{m_1}, \ldots, \bar{m_k})$.\label{def-rep-rel-cond-ii}
    \end{enumerate}
\end{definition}

\begin{definition}\label{def-rep-comp}
    Let $\mc{M}$ be a TM, $\alpha$ the input for $\mc{M}$, and
     $\mu(\mc{M}(\alpha)) = \langle \mbb{Z}, Q_1^{\mu}, Q_2^{\mu}, \dots, Q_n^{\mu}, S_1^{\mu}, S_1^{\mu}, \dots, S_{m}^{\mu}, <^{\mu}, '^{\mu}, 0^{\mu} \rangle$
     the structure determined by the intended interpretation $\mc{I}$.\footnote{
     $\mbb{Z}$ represents the integers, the
     relations  $Q_i^{\mu}$  express couples of instants of time and positions for
     states $q_i$ in the computation of $\mc{M}(\alpha)$, relations $S_j^{\mu}$
     express couples of instants of time and positions for symbols $s_j$ in the computation
     of $\mc{M}(\alpha)$,  $<^{\mu}$ is the standard strict order on $\mbb{Z}$, $'^{\mu}$
     is the successor function on $\mbb{Z}$ and $0^{\mu}$ is the integer $0$.}.
     A theory $\Delta$, in the language $\mc{L} = \{Q_1, Q_2, \ldots, Q_n, S_0, S_1, \ldots, S_{m-1}, <, ', 0\}$,
     \emph{represents the computation of $\mc{M}(\alpha)$} if:
    \begin{enumerate}
        \item $<^{\mu}$ is represented by $\varphi(x, y) := x < y$ in $\Delta$,
        \item $'^{\mu}$ is represented by $\varphi(x, y) := x' = y$ in $\Delta$,
        \item $Q_i^{\mu}$ ($i = 1, \ldots, n$) are represented by $\varphi(x, y) := Q_i(x, y)$ in $\Delta$, and
        \item $S_j^{\mu}$ ($j = 1, \ldots, m$) are represented by $\varphi(x, y) := S_j(x, y)$ in $\Delta$.
    \end{enumerate}
\end{definition}

\begin{theorem}\label{theo-delta-not-rep}
Let $\mc{M}$ be a TM and $\alpha$ the input for $\mc{M}$. The
theory $\Delta_{FOL}(\mc{M}(\alpha))$ cannot represent the
computation of $\mc{M}(\alpha)$.
\begin{proof}
    We show  that condition \ref{def-rep-rel-cond-ii} of Definition \ref{def-rep-rel}
    cannot be satisfied for relations $Q_i$ and $S_j$: Indeed, when $\mc{M}(\alpha)$ is in
    state $q_i$, at time $t$ and position $x$, it is not in any other state
    $q_j$ ($i \neq j$); in this case, we have
    that $\Delta_{FOL}(\mc{M}(\alpha)) \vdash Q_i(\bar{t}, \bar{x})$
    (by the proof in \cite[Chap. 10]{Boolos-Jeffrey-1989}),
    but on the other hand  we have
    that $\Delta_{FOL}(\mc{M}(\alpha)) \nvdash \neg Q_j(\bar{t}, \bar{x})$,
    because a non-standard TM with the same instructions of $\mc{M}$ (but
    allowing multiple simultaneous states: starting the computation in
    two-different simultaneous states, for example) also validates all
    axioms in $\Delta_{FOL}(\mc{M}(\alpha))$. A similar situation occurs
    with relations $S_j$. We can also define other non-standard TMs which
    allow different symbols and states, on different positions of the tape,
    at times before the beginning or after the end of the computation,
    in such a way that the machine validates all axioms in $\Delta_{FOL}(\mc{M}(\alpha))$.
\end{proof}
\end{theorem}

Theorem \ref{theo-delta-not-rep} shows that it is necessary to
expand the theories $\Delta_{FOL}(\mc{M}(\alpha))$ in order to disallow
non-standard interpretations and to grant  representation of
computations in accordance with Definition \ref{def-rep-comp}. We thus
define the  notion  of an \emph{intrinsic  theory of the
computation of} $\mc{M}(\alpha)$ as the theory
$\Delta^{\star}_{FOL}(\mc{M}(\alpha))$ by specifying which new
axioms have to be added to $\Delta_{FOL}(\mc{M}(\alpha))$
theories, so that these extended theories  are able to  represent
their respective TM computations (Theorem
\ref{theo-delta-exp-rep}). For the specification of such axioms,
we will suppose that before the beginning of any computation and
after the end of any computation (if the computation halts), the
machine is in none of its states and no symbols (not even the
empty symbol) occurs anywhere in its tape. New axioms are defined
as follows:
\begin{itemize}
    \item An axiom to define the situation of $\mc{M}(\alpha)$ before the beginning of
    the computation:
        \begin{equation}
            \forall t \forall x \left((t < 0) \to \left(\left(\bigwedge_{i=1}^{n} \neg Q_i(t, x)\right) \wedge \left(\bigwedge_{j=1}^{m} \neg S_j(t, x)\right)\right)\right).\tag{A$t0$} \label{ax-t-0}
        \end{equation}

    \item An axiom to define the situation of $\mc{M}(\alpha)$ after the
     end of the computation (if the computation halts):
        \begin{multline}
            \forall t \forall x \Biggr(\neg \Biggl(\bigvee_{q_i s_j \in I} \biggl(Q_i(t, x) \wedge S_j(t, x)\biggr)\Biggr) \to \\
            \forall u \forall y \Biggl( t < u \to \biggl(\biggl(\bigwedge_{i=1}^{n}\neg Q_i(u, y) \biggr) \wedge \biggl(\bigwedge_{j=1}^{m}\neg S_j(u, y)\biggr)\biggr)\Biggr)\Biggr), \tag{A$th$} \label{ax-t-halt}
        \end{multline}
        where subscript $q_i s_j \in I$ means that, in the disjunction, only
         combinations of $q_i s_j$ coincident with the  first two symbols of some instruction
         of $\mc{M}$  are taken into account.

    \item An axiom for any state symbol $q_i$ of $\mc{M}$
    establishing the uniqueness  of any  state and  any position in a given  instant of time:
        \begin{equation}
            \forall t \forall x \left(Q_i(t, x) \to \left(\left(\bigwedge_{j \neq i} \neg Q_j(t, x)\right) \wedge \forall y \left(y \neq x \to \bigwedge_{i=1}^{n}\neg Q_i(t, y)\right)\right)\right). \tag{A$q_i$} \label{ax-unity-state}
        \end{equation}

    \item An axiom for any read/write symbol $s_i$ of $\mc{M}$
    establishing the uniqueness  of any symbol in a given instant of time and position:
        \begin{equation}
            \forall t \forall x \left(S_i(t, x) \to \bigwedge_{i \neq j} \neg S_j(t, x)\right).\tag{A$s_j$}\label{ax-unity-symbol}
        \end{equation}

\end{itemize}

\begin{theorem}\label{theo-delta-exp-rep}
Let $\mc{M}$ be a TM and $\alpha$ the input for $\mc{M}$. Then, the
intrinsic theory $\Delta^{\star}_{FOL}(\mc{M}(\alpha))$ represents
the computation of $\mc{M}(\alpha)$.
\begin{proof}
Representation for the  relation $<^{\mu}$ and for the  function
$'^{\mu}$ is  easy to proof. Representation for  relations
$Q_i^{\mu}$ and $S_j^{\mu}$ follows from the proof in \cite[Chap.
10]{Boolos-Jeffrey-1989} and direct applications of the new
axioms in the intrinsic theory
$\Delta^{\star}_{FOL}(\mc{M}(\alpha))$.
\end{proof}
\end{theorem}

The definitions and theorems above consider only
DTMs (i.e, TMs with no pairs of instructions with the same two initial symbols);
the next theorem establishes that the method of axiomatization
defined above, when used to NDTMs, produces
contradictory theories (in some cases).

\begin{theorem}\label{theo-NDTM-cont}
Let $\mc{M}$ be a NDTM and $\alpha$ an  input for $\mc{M}$. If
$\mc{M}(\alpha)$ reaches an ambiguous configuration (i.e. a configuration where multiple instructions can be executed), then its
intrinsic theory  $\Delta^{\star}_{FOL}(\mc{M}(\alpha))$ is
contradictory.
\begin{proof}
By the proof in \cite[Chap. 10]{Boolos-Jeffrey-1989}, it is deduced a formula that expresses the ambiguous configuration. Then, by using theorems corresponding to the possible instructions that can be executed in the ambiguous configuration, there are deduced formulas expressing multiplicity of states, positions or symbols in some cell of the tape. Thus, by using axiom \eqref{ax-unity-state} or \eqref{ax-unity-symbol}, a contradiction is deduced.
\end{proof}
\end{theorem}

In \cite[p. 48]{Odifreddi-1989}, Odifreddi, in his definition of a
TM, establishes a condition of ``consistency'' for the machine
disallowing the existence of ``contradictory'' instructions (i.e.,
instructions with the same two initial symbols), which
corresponds to the notion of DTM. Thus, NDTMs are those that do not accomplish
the condition of consistency. Theorem
\ref{theo-NDTM-cont} shows that Odifreddi's idea of consistency in
TMs coincides with the consistency of the intrinsic theories
$\Delta^{\star}_{FOL}(\mc{M}(\alpha))$.

Note that contradictions in intrinsic theories
$\Delta^{\star}_{FOL}(\mc{M}(\alpha))$  arise by the multiple use
of axioms \eqref{ax-inst-i}, \eqref{ax-inst-ii} and
\eqref{ax-inst-iii} for the same instance of $t$ in combination
with the use of axioms \eqref{ax-unity-state} or
\eqref{ax-unity-symbol}. The multiple use of axioms
\eqref{ax-inst-i}, \eqref{ax-inst-ii} and \eqref{ax-inst-iii} for
the same instance of $t$ indicates the simultaneous execution of
multiple instructions, which can derive (at time $t + 1$)
multiplicity of symbols in the cell of the tape where the
instruction is executed, or multiplicity of states and positions,
while axioms \eqref{ax-unity-state} and \eqref{ax-unity-symbol}
establish the uniqueness  of such  elements. Intrinsic theories
$\Delta^{\star}_{FOL}(\mc{M}(\alpha))$ can be easily adapted to
deal with the idea that  only one instruction  is chosen to be
executed when the machine reaches an ambiguous configuration,
obtaining adequate theories for NDTMs computations. However, this
is not our focus in this paper. We are interested in
generalizing the notion of TM by using a paraconsistent logic, as
this is a fecund way of approaching quantum computing from a
logical viewpoint.

\section{Paraconsistent TMs}\label{ptms}

There are many paraconsistent logics. They are proposed from different
philosophical perspectives but share the common feature of being
logics which support contradictions without falling into deductive
trivialization. Although  in the definition of ParTMs we could, in
principle, depart from  any first-order paraconsistent logic, we
will use the logic $LFI1^*$ (see
\cite{Carnielli-Marcos-deAmo-2000}), because it possesses an already
established proof-theory and first-order semantics, has properties
that allows natural interpretations of consequences of
$\Delta^{\star}_{LFI1^*}(\mc{M}(\alpha))$
theories\footnote{Intrinsic theories
$\Delta^{\star}_{LFI1^*}(\mc{M}(\alpha))$  are obtained
 by substituting the underlying logic of
$\Delta^{\star}_{FOL}(\mc{M}(\alpha))$ theories by $LFI1^*$.} as
`paraconsistent computations', and also allows the addition of
conditions to control the execution of instructions involving
multiplicity of symbols and states.\footnote{It is worth to
remark that the choice of another paraconsistent logic, with
other features, can lead to different notions of ParTMs, as is
the case in Sec. \ref{sim-ent-states-rel-phases}.} $LFI1^*$ is the
first-order extension of $LFI1$, which is an LFI that extends
positive classical logic, defines connectives of
consistency $\circ$ and inconsistency $\bullet$,
and identifies inconsistency with contradiction by  means of the
equivalence $\bullet A \leftrightarrow (A \wedge \neg A)$.

For intrinsic theories $\Delta^{\star}_{LFI1^*}(\mc{M}(\alpha))$
the proof in \cite[Chap. 10]{Boolos-Jeffrey-1989} continues to
hold, because $LFI1^*$ is an extension of positive classical
logic. Thus, as described above, the  use of multiple axioms
describing instructions for the same instance of $t$ indicates
simultaneous execution of the instructions, which gives place to
multiplicity of symbols in the cell and multiplicity of states
and positions. Such a  multiplicity, in conjunction with axioms
\eqref{ax-unity-state} and \eqref{ax-unity-symbol}, entails
contradictions  which are identified in $LFI1^*$ with
inconsistencies. Thus, inconsistency in $\Delta^{\star}_{LFI1^*}(\mc{M}(\alpha))$ theories
characterize multiplicity.

By taking advantage of the robustness of $LFI1^*$  in the presence
of inconsistencies and their interpretation as multiplicity, we
can supply the ParTMs with conditions of inconsistency in the
two initial symbols of instructions in order to control the
process of computation. $q_i^{\bullet}$ will indicate that the
instruction  will  only be executed in configurations where the
machine is in multiple states or multiple positions, and
$s_j^{\bullet}$ will indicate that the instruction will only be
executed  in cells with multiple symbols. These conditions
correspond to put the connective $\bullet$, respectively,  in
front of  the predicate $Q_i$ or $S_j$ in the antecedent of the axioms
related to the instructions. These apparently innocuous conditions are
essential for taking advantage of the parallelism provided by ParTMs and EParTMs.
As will be argued below, inconsistency conditions on the instructions seem to be
a more powerful mechanism than quantum interference, which is the instrument provided
by quantum computation taking advantage of quantum parallelism.

Note that axioms \eqref{ax-inst-i}, \eqref{ax-inst-ii} and
\eqref{ax-inst-iii} not only express the action of instructions
but also specify the preservation of symbols unmodified by the
instructions. Thus, in ParTMs, we have to take into account that any
instruction is executed in a specific position on the tape,
carrying  symbols from cells not modified by the instruction to the next
instant of time; this is completed independently of the execution of other
instructions.

A ParTM is then defined as:
\begin{definition}
A \emph{ParTM} is a NDTM such that:
\begin{itemize}
    \item When the machine reaches an ambiguous configuration it \emph{simultaneously} executes
    all possible instructions, which can produce multiplicity on states, positions and symbols
     in some cells of the tape;
    \item Each instruction is executed in the position corresponding to the respective
    state; symbols in cells unmodified by the instructions are carried  to the next instant
    of time;
    \item \emph{Inconsistency} (or \emph{multiplicity}) conditions are allowed on the first
     two symbols of the instructions (as described above);
    \item The machine stops when there are no instructions to be executed; at this stage some
    cells of the tape can contain multiple symbols, any choice of them represents a result of
     the computation.
\end{itemize}
\end{definition}

The next example illustrates how a ParTM performs computations:
\begin{example}\label{exam-ParTM}
Let $\mc{M}$ be a ParTM with instructions: $i_1: q_1 0 0 q_2$,
$i_2: q_1 0 1 q_2$, $i_3: q_2 0 R q_3$, $i_4: q_2 1 R q_3$, $i_5:
q_3 \emptyset 1 q_4$, $i_6: q_4 0 0 q_5$, $i_7: q_4 1 0 q_5$,
$i_8: q_4 1^{\bullet} * q_5$, $i_9: q_5 * 1 q_5$. Figure
\ref{fig-comp-ParTM} schematizes the computation of $\mc{M}$,
beginning in position $0$, state $q_1$ and reading the symbol $0$
(with symbol $\emptyset$ in all other cells of the tape).
Instructions to be executed in each instant of time $t$ are
written within  parentheses (note that instruction $i_8$ is not
executed at time $t = 3$ because of the inconsistency condition
on the scanned symbol). $\mc{M}$ will
be useful in the paraconsistent solution of Deutsch's and
Deutsch-Jozsa problems (Sec. \ref{sim-D-DJ-prob}).
\begin{figure}[ht]
\psset{unit=4mm}
\begin{center}
   \scriptsize
  \begin{picture}(125, 140)
  %\thicklines

  \put(0, 140){$t = 0 \; (i_1, i_2)$}
  \put(25, 130){\ldots}
  \put(40, 125){\framebox(15, 10){$\emptyset$}}
  \put(55, 125){\framebox(15, 10){$\emptyset$}}
  \put(70, 125){\framebox(15, 10){$0$}}
  \put(85, 125){\framebox(15, 10){$\emptyset$}}
  \put(100, 125){\framebox(15, 10){$\emptyset$}}
  \put(118, 130){\ldots}
  \put(40, 120){$_{-2}$}
  \put(55, 120){$_{-1}$}
  \put(75, 120){$_{0}$}
  \put(90, 120){$_{1}$}
  \put(105, 120){$_{2}$}
  \put(75, 145){\vector(0, -1){10}}
  \put(78, 140){$q_1$}

  \put(0, 110){$t = 1 \; (i_3, i_4)$}
  \put(25, 100){\ldots}
  \put(40, 95){\framebox(15, 10){$\emptyset$}}
  \put(55, 95){\framebox(15, 10){$\emptyset$}}
  \put(70, 95){\framebox(15, 10){$0, 1$}}
  \put(85, 95){\framebox(15, 10){$\emptyset$}}
  \put(100, 95){\framebox(15, 10){$\emptyset$}}
  \put(118, 100){\ldots}
  \put(40, 90){$_{-2}$}
  \put(55, 90){$_{-1}$}
  \put(75, 90){$_{0}$}
  \put(90, 90){$_{1}$}
  \put(105, 90){$_{2}$}
  \put(75, 115){\vector(0, -1){10}}
  \put(78, 110){$q_2$}

  \put(0, 80){$t = 2 \; (i_5)$}
  \put(25, 70){\ldots}
  \put(40, 65){\framebox(15, 10){$\emptyset$}}
  \put(55, 65){\framebox(15, 10){$\emptyset$}}
  \put(70, 65){\framebox(15, 10){$0, 1$}}
  \put(85, 65){\framebox(15, 10){$\emptyset$}}
  \put(100, 65){\framebox(15, 10){$\emptyset$}}
  \put(118, 70){\ldots}
  \put(40, 60){$_{-2}$}
  \put(55, 60){$_{-1}$}
  \put(75, 60){$_{0}$}
  \put(90, 60){$_{1}$}
  \put(105, 60){$_{2}$}
  \put(90, 85){\vector(0, -1){10}}
  \put(93, 80){$q_3$}

  \put(0, 50){$t = 3 \; (i_7)$}
  \put(25, 40){\ldots}
  \put(40, 35){\framebox(15, 10){$\emptyset$}}
  \put(55, 35){\framebox(15, 10){$\emptyset$}}
  \put(70, 35){\framebox(15, 10){$0, 1$}}
  \put(85, 35){\framebox(15, 10){$1$}}
  \put(100, 35){\framebox(15, 10){$\emptyset$}}
  \put(118, 40){\ldots}
  \put(40, 30){$_{-2}$}
  \put(55, 30){$_{-1}$}
  \put(75, 30){$_{0}$}
  \put(90, 30){$_{1}$}
  \put(105, 30){$_{2}$}
  \put(90, 55){\vector(0, -1){10}}
  \put(93, 50){$q_4$}

  \put(0, 20){$t = 4$}
  \put(25, 10){\ldots}
  \put(40, 5){\framebox(15, 10){$\emptyset$}}
  \put(55, 5){\framebox(15, 10){$\emptyset$}}
  \put(70, 5){\framebox(15, 10){$0, 1$}}
  \put(85, 5){\framebox(15, 10){$0$}}
  \put(100, 5){\framebox(15, 10){$\emptyset$}}
  \put(118, 10){\ldots}
  \put(40, 0){$_{-2}$}
  \put(55, 0){$_{-1}$}
  \put(75, 0){$_{0}$}
  \put(90, 0){$_{1}$}
  \put(105, 0){$_{2}$}
  \put(90, 25){\vector(0, -1){10}}
  \put(93, 20){$q_5$}

  \end{picture}
\end{center}
\caption{\scriptsize Example of computation in a ParTM}
\label{fig-comp-ParTM}
\end{figure}
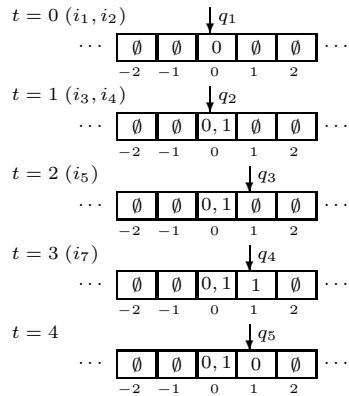
\end{example}

\subsection{Simulating Quantum Computation through Paraconsistent TMs}\label{sim-qc-ptms}

In the Neumann-Dirac formulation, quantum mechanics is
synthesized in four postulates (cf. \cite[Sec.
2.2]{Nielsen-Chuang-2000}): The first postulate establishes that
states of isolated physical systems are represented by unit
vectors in a Hilbert space (known as the \emph{space state} of
the system); the second postulate claims that the evolution of closed
quantum systems is described by unitary transformations in the
Hilbert space; the third postulate deals with observations of
physical properties of the system by relating physical properties
with Hermitian operators (called \emph{observables}) and
establishing that, when a measurement is performed, an eigenvalue
of the observable is obtained (with a certain probability
depending upon the state of the system) and the system collapses to
its respective eigenstate; finally, the fourth postulate
establishes  the tensor product of the state spaces of the component
systems as being the state space of the compound system,
allowing us to represent the state of a compound system as the tensor
product of the state of its subsystems (when the states of the
subsystems are known).

The best known models of quantum computation, namely \emph{quantum
Turing machines} (QTMs)  and \emph{quantum circuits} (QCs), are
direct generalizations of TMs and boolean
circuits, respectively, using the laws of quantum mechanics.

By taking into account the postulates of quantum mechanics briefly
described above, a QTM (introduced in \cite{Deutsch-1985}) is
defined by considering elements of a TM (state, position and
symbols on the tape) as being observables of a quantum
system. The configuration of a QTM is thus represented by a unit vector in a Hilbert
space and the evolution is described by a unitary operator (with
some restrictions in order to satisfy the requirement that the machine
operates by `finite means', see \cite[p. 7]{Deutsch-1985} and
\cite{Ozawa-Nishimura-2000}). Since the configuration of a QTM
is described by a unit vector, it is generally a linear
superposition of basis states (called a \emph{superposed
state}), where the basis states represent classical TM
configurations. In quantum mechanics, superposed states can be
interpreted as the coexistence of multiple states, thus a QTM
configuration can be interpreted as the simultaneous existence of
multiple classical TM configurations. The linearity of the
operator describing the evolution of a QMT allows us to think in
the parallel execution of the instructions over the different states
(possibly an exponential number) present in the superposition.
Unfortunately, to know the result of the computation, we have to
perform a measurement of the system and, by the third postulate
of quantum mechanics, we can obtain  only one classical TM
configuration in a probabilistic way, in effect, irredeemably losing all
other configurations. The art of `quantum programming' consists
in taking advantage of the intrinsic parallelism of the model,
by using quantum interference\footnote{Quantum interference is
expressed by the addition of amplitudes corresponding to equal basis states in a superposed state.
When signs of amplitudes are the same, their sum obtains a greater amplitude; in this case, we say that the interference is \emph{constructive}.
Otherwise, amplitudes subtract and we say that the interference is \emph{destructive}. Quantum interference occurs in the evolution of one quantum state to another.} to increase amplitudes of desired states
before the measurement of the system, with the aim to solve problems more
efficiently than in the classical case.

As shown in \cite{Ozawa-Nishimura-2000}, the evolution of a QTM
can be equivalently specified by a \emph{local transition
function}\footnote{Some changes were made in the definition of
$\delta$ to deal with the quadruple notation for instructions we
are using here.} $\fp{\delta} Q \times \Sigma \times \{\Sigma \cup
\{R, L\}\} \times Q \to \mbb{C}$, which validates some conditions related
to the unitary of state vectors and the reversibility of operators. In
this definition, the transition $\delta(q_i, s_j, Op, q_l) = c$
can be interpreted as the following action of the QTM: If the
machine is in state $q_i$ reading the symbol $s_j$, it follows
with the probability amplitude $c$ that the machine will perform  the
operation $Op$ (which can be either to write a symbol or to move
on the tape) and reaches the state $q_l$. The amplitude $c$ cannot
be interpreted as the probability of performing the respective
transition, as with probabilistic TMs. Indeed, QTMs do not choose
only one transition to be executed; they can perform multiple
transitions simultaneously in a single instant of time in
a superposed configuration. Moreover, configurations resulting from different
transitions can interfere constructively or destructively (if they represent the same classical configuration), respectively increasing
or decreasing the amplitude of the configuration in the superposition.

By taking into account that each choice function on the elements of
a ParTM, in an instant of time $t$, gives a classical TM
configuration; a configuration of a ParTM can be viewed as a
\emph{uniform}\footnote{A superposed state is said to be \emph{uniform} if
all states in the superposition, with amplitude different of $0$,
have amplitudes with the same magnitude.}
superposition of classical TM configurations. This way, ParTMs
seem to be similar to QTMs: We could see ParTMs as QTMs without
amplitudes (which allows us only to represent uniform superpositions).
However, in ParTMs, actions performed by different instructions
mix  indiscriminately, and  thus  all combinations of the singular
elements in a ParTM configuration are taken into account, which
makes it impossible to represent entangled states by only considering the
multiplicity of elements as superposed states (this point is
discussed in Sec. \ref{sim-ent-states-rel-phases}). Another difference
between the  ParTMs and QTMs models is that `superposed states'
in the former model do  not supply  a notion of relative phase (corresponding to signs of basis states in uniform superpositions),
an important feature of quantum superpositions required for quantum interference.
Such a feature, as mentioned before, is the key mechanism  for taking advantage of quantum parallelism.
However, inconsistency conditions on the instructions of ParTMs allows us to take
advantage of `paraconsistent parallelism', and this seems to be a
more powerful property  than quantum interference (this point is fully discussed in
Sec. \ref{sim-ent-states-rel-phases}). In spite of the differences
between ParTMs and QTMs, ParTMs are able to simulate important
features of quantum computing; in particular, they can simulate uniform non-entangled superposed quantum states and solve the Deutsch and Deutsch-Jozsa problems preserving the efficiency  of the quantum algorithms, but with  certain restrictions  (see Sec. \ref{sim-D-DJ-prob}). In Sec.
\ref{sim-ent-states-rel-phases}, we define another model of ParTMs, based on
a paraconsistent logic endowed with a `non-separable' conjunction, which
enables the simulation of uniform entangled states and represents a
better approach for the model of QTMs. We also show that  a notion
of `relative phase' can be introduced in this new model of computation.

\subsubsection{Paraconsistent Solutions for Deutsch and Deutsch-Jozsa Problems}\label{sim-D-DJ-prob}

Given an arbitrary function $\fp{f} \{0, 1\} \to \{0, 1\}$ and an
`oracle' (or black box) that computes $f$, Deutsch's problem
consists in defining a procedure to determine if $f$ is
\emph{constant} ($f(0) = f(1)$) or \emph{balanced} ($f(0) \neq
f(1)$) allowing only one query to the oracle. Classically, the procedure
seems to require two queries to the oracle in order  to compute $f(0)$
and $f(1)$, plus a  further step  for the comparison; but by taking
advantage of the quantum laws the problem can be solved in a more efficient way,
by executing just a single query.
A probabilistic quantum solution to Deutsch's problem was first
proposed in \cite{Deutsch-1985} and a deterministic quantum
algorithm was given in
\cite{Cleve-Ekert-Macchiavello-Mosca-1998}. The deterministic
solution is usually formulated  in the QCs formalism, so we
briefly describe this model of computation before presenting the
quantum algorithm.

\begin{figure}[ht]
  \psset{unit=5mm}
  \centering
  \begin{pspicture}(0,-2)(13,4)%\grilla
    \rput[c](2.5,-1){$\ket{\psi_0}$}
    \psline{->}(2.5,-0.5)(2.5,0.5)
    \rput[c](6,-1){$\ket{\psi_1}$}
    \psline{->}(6,-0.5)(6,0.5)
    \rput[c](9.5,-1){$\ket{\psi_2}$}
    \psline{->}(9.5,-0.5)(9.5,0.5)
    \rput[c](12.5,-1){$\ket{\psi_3}$}
    \psline{->}(12.5,-0.5)(12.5,0.5)

    \uput[l](2,1){$\ket{1}$}
    \psline(2,1)(3.2,1)
    \psframe[linewidth=1.5pt,framearc=0.2](3.2,0.2)(4.8,1.8)
    \rput[c](4,1){$H$}
    \psline(4.8,1)(7,1)
    \psframe[linewidth=1.5pt,framearc=0.2](7,0)(9,4)
    \rput[c](8,2){$U_f$}
    \psline(9,1)(13,1)

    \uput[l](2,3){$\ket{0}$}
    \psline(2,3)(3.2,3)
    \psframe[linewidth=1.5pt,framearc=0.2](3.2,2.2)(4.8,3.8)
    \rput[c](4,3){$H$}
    \psline(4.8,3)(7,3)
    \psline(9,3)(10.2,3)
    \psframe[linewidth=1.5pt,framearc=0.2](10.2,2.2)(11.8,3.8)
    \rput[c](11,3){$H$}
    \psline(11.8,3)(13,3)
  \end{pspicture}
  \caption{\scriptsize QC to solve Deutsch's problem}
  \label{fig-qc-Deutsch-problem}
\end{figure}
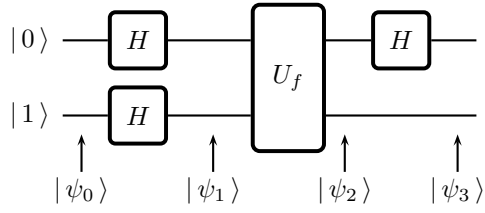

The model of QCs (introduced in \cite{Deutsch-1989}) is defined
by generalizing the  boolean circuit model in accordance with
the postulates of quantum mechanics: The classical unit of
information, the \emph{bit}, is generalized as the \emph{quantum
bit} (or \emph{qubit}), which is mathematically represented by a
unit vector in a two-dimensional Hilbert space; classical
logic gates are replaced by unitary operators; registers of
qubits are represented by tensor products and measurements
(following conditions of the third postulate above) are accomplished
at the end of the circuits in order to obtain the output of the
computation.\footnote{For a detailed introduction to QCs see
\cite{Nielsen-Chuang-2000}.} Under this definition, the QC
depicted in Figure \ref{fig-qc-Deutsch-problem} represents a
deterministic solution to Deutsch's problem.

In the figure, squares labeled by $H$ represent \emph{Hadamard}
gates. A Hadamard gate is a quantum gate which performs the
following transformations ($\ket{\cdot}$ representing a vector in
Dirac's notation):
\begin{align}
  \fp{H} &\ket{0} \mapsto \frac{1}{\sqrt{2}} \left(\ket{0} + \ket{1}\right) \nonumber \\
  &\ket{1} \mapsto \frac{1}{\sqrt{2}} \left(\ket{0} - \ket{1}\right).
\end{align}
The rectangle labeled by $U_f$ represents the quantum oracle that
performs the operation $U_f(\ket{x, y}) = \ket{x, y \oplus
f(x)}$, where $\ket{x, y}$ represents the tensor product
$\ket{x} \otimes \ket{y}$ and $\oplus$ represents the addition module
2. Vectors $\ket{\psi_i}$ are depicted to explain, step by step, the process
of computation:
\begin{enumerate}
    \item At the beginning of the computation, the input register takes the value $\ket{\psi_0} = \ket{0, 1}$;
    \item After performing the two first Hadamard gates, the following superposition
    is obtained:
        \begin{equation}
        \ket{\psi_1} = H \ket{0} \otimes H \ket{1} = \frac{1}{2}\left((\ket{0} + \ket{1}) \otimes (\ket{0} - \ket{1})\right);\label{eq-state1-qc-dp}
        \end{equation}
    \item By applying the operation $U_f$, one obtains:
        \begin{align}
        \ket{\psi_2} &= U_f \left(\frac{1}{2}\left(\ket{0, 0} - \ket{0, 1} + \ket{1, 0} - \ket{1, 1}\right)\right)  \label{eq-state2-qc-dp} \\
        &= \frac{1}{2}\left(\ket{0, 0 \oplus f(0)} - \ket{0, 1 \oplus f(0)} + \ket{1, 0 \oplus f(1)} - \ket{1, 1 \oplus f(1)}\right) \nonumber \\
        &= \frac{1}{2}\left((-1)^{f(0)} (\ket{0} \otimes (\ket{0} - \ket{1})) + (-1)^{f(1)} (\ket{1} \otimes (\ket{0} - \ket{1}))\right) \nonumber \\
        &=
            \begin{cases}
            \pm \left(\frac{1}{\sqrt{2}}(\ket{0} + \ket{1})\right) \otimes \left(\frac{1}{\sqrt{2}}(\ket{0} - \ket{1})\right) \mbox{ if } f(0) = f(1),\\
            \pm \left(\frac{1}{\sqrt{2}}(\ket{0} - \ket{1})\right) \otimes \left(\frac{1}{\sqrt{2}}(\ket{0} - \ket{1})\right) \mbox{ if } f(0) \neq f(1).
            \end{cases} \nonumber
        \end{align}
    \item By applying the last Hadamard gate, one finally reaches:
        \begin{equation}
        \ket{\psi_3} =
            \begin{cases}
            \pm \ket{0} \otimes \left(\frac{1}{\sqrt{2}}(\ket{0} - \ket{1})\right) \mbox{ if } f(0) = f(1),\\
            \pm \ket{1} \otimes \left(\frac{1}{\sqrt{2}}(\ket{0} - \ket{1})\right) \mbox{ if } f(0) \neq f(1).
            \end{cases}\label{eq-state3-qc-dp}
            \end{equation}
\end{enumerate}
After a measurement of the first qubit of  the state $\ket{\psi_3}$
is accomplished (on the standard basis, cf. \cite{Nielsen-Chuang-2000}),
one obtains  $0$ (with probability $1$) if $f$ is constant or $1$
(with probability $1$) if $f$ is balanced.

The first step of the above QC  generates a superposed state
(Eq. \eqref{eq-state1-qc-dp}), which is taken into the next step
to compute the function $f$ in parallel  (Eq.
\eqref{eq-state2-qc-dp}), generating a superposition in such a way that
the relative phase of $\ket{1}$ in the first qubit differs depending on if $f$ is constant or balanced.
By applying again a Hadamard gate on the first qubit, quantum interference acts leaving the first qubit in the
basis state $\ket{0}$ if $f$ is constant or in the basis state $\ket{1}$ if $f$ is balanced (Eq. \eqref{eq-state3-qc-dp}). Thus,
by performing a measurement of the first qubit, we determine
with certainty if $f$ is constant or balanced. Note that
$U_f$ is used only once in the computation.

The ParTM in Example \ref{exam-ParTM} gives a `paraconsistent' simulation of
the quantum algorithm that  solves Deutsch's problem, for the
particular case where $f$ is the constant function $1$.
Instructions $i_1$ and $i_2$, executed simultaneously at time $t
= 0$, simulate the generation of the superposed state.
Instructions $i_3$ to $i_5$ compute  the constant function $1$
over the superposed state, performing in parallel the computation of
$f(0)$ and $f(1)$, and writing the results on position $1$ of the
tape. Instructions $i_6$ to $i_9$ check whether  $f(0) = f(1)$, writing
$0$ on position $1$ of the tape if there is no multiplicity of
symbols on the cell (meaning that $f$ is constant) or writing $1$
in another case (meaning that $f$ is balanced). In the present case
$f(0) = f(1) = 1$, thus the execution of the instructions from $i_6$ to $i_9$
gives as result the writing of $0$ on position $1$ of the tape.

Consider a TM $\mc{M}'$ a black box that computes a function
$\fp{f} \{0, 1\} \to \{0, 1\}$. We could substitute instructions
$i_3$ to $i_5$ in Example \ref{exam-ParTM} (adequately renumbering
instructions and states from $i_6$ to $i_9$ if necessary) with the
instructions from $\mc{M}'$ in order to determine if $f$ is constant or
balanced. In this way, we define a paraconsistent simulation of the
quantum algorithm that  solves Deutsch's problem. In the simulation,
$\mc{M}'$ is the analog of $U_f$ and quantum parallelism is
mimicked by the parallelism provided by the multiplicity
allowed in the ParTMs.

Notwithstanding the parallelism provided
by ParTMs, this first paraconsistent model of computation has some peculiar
properties which could give rise to `anomalies' in the process of
computation. For instance, consider a TM $\mc{M}'$ with
instructions: $i_1 = q_1 0 0 q_2$, $i_2 = q_1 1 1 q_3$, $i_3 =
q_2 0 R q_4$, $i_4 = q_3 1 R q_4$, $i_5 = q_2 1 R q_5$, $i_6 =
q_4 \emptyset 1 q_4$, $i_7 = q_5 \emptyset 0 q_5$. If $\mc{M}'$
starts the computation on position $0$ and state $q_1$, with a
single symbol ($0$ or $1$) on position $0$ of the tape, and with
symbol $\emptyset$ on all other cells of the tape, then $\mc{M}'$
computes the constant function $1$. Nevertheless, if $\mc{M}'$
begins reading symbols $0$ and $1$ (both simultaneously on cell
$0$), then it produces $0$ and $1$ as its outputs, as if $\mc{M}'$
had computed a balanced function. This example shows well how different
paths of a computation can mix indiscriminately, producing paths of computation
not possible in the TM considered as an oracle (when viewed as a classical TM) and generating undesirable results.
Therefore, only TMs that exclusively perform their possible paths of computation
when executed on a superposed state can be considered as oracles.
TMs with this property will be called \emph{parallelizable}.
 Note that this restriction is not a serious  limitation to our paraconsistent
solution of Deutsch's problem, because parallelizable TMs that compute any
function $\fp{f} \{0, 1\} \to \{0, 1\}$ are easy to define.

The Deutsch-Jozsa problem, first presented in
\cite{Deutsch-Jozsa-1992}, is the generalization of Deutsch's
problem to functions $\fp{f} \{0, 1\}^n \to \{0, 1\}$, where $f$
is assumed  to be either  constant or balanced.\footnote{$f$ is balanced
if $\card{f^{-1}(0)} = \card{f^{-1}(1)}$, where $\card{A}$
represents the cardinal of the set $A$.} A quantum solution to the
Deutsch-Jozsa problem is a direct generalization of the quantum
solution to Deutsch's problem presented above: The input register
is now constituted of $n + 1$ qubits and takes the value
$\ket{0}^{\otimes n} \otimes \ket{1}$ (where
$\ket{\cdot}^{\otimes n}$ represents the tensor product of $n$
qubits $\ket{\cdot}$); new Hadamard gates are added to act on the
new qubits in the input register and also on the first $n$
outputs of $U_f$, and $U_f$ is now a black box acting on $n + 1$
qubits, performing the operation $U_f(\ket{x_1, \ldots, x_n, y})
= \ket{x_1, \ldots, x_n, y \oplus f(x_1, \ldots, x_n)}$. In this
case, when a measurement of the first $n$ qubits is accomplished
at the end of the computation, if all the values obtained are $0$,
then $f$ is constant (with probability $1$); in another case $f$ is
balanced (with probability $1$).\footnote{Calculations are not
presented here, for details see \cite{Nielsen-Chuang-2000}.}

The paraconsistent solution to Deutsch's problem can be easily
generalized to solve the Deutsch-Jozsa problem as well: The input to
$\mc{M}$ must be a sequence of $n$ symbols $0$; instructions
$i_1$ and $i_2$ must be substituted by instructions $i_1 = q_1 0
0 q_2$, $i_2 = q_1 0 1 q_2$, $i_3 = q_1 0 R q_1$, $i_4 = q_1
\emptyset L q_3$, $i_5 = q_3 0 L q_3$, $i_6 = q_3 \emptyset R
q_4$,  and the machine $\mc{M}'$ is now considered to be a parallelizable TM computing a
constant or balanced function $\fp{f} \{0, 1\}^n \to \{0, 1\}$.

Note that the paraconsistent solution to the Deutsch-Jozsa problem does not depend on
the assumption of the function to be constant or balanced; in fact, the solution can be
applied in order to distinguish between constant and non-constant functions.
This can lead to the erroneous conclusion that PartTMs are too powerful machines, in which all NP-Problems could  be solved in polynomial time: Someone could  mistakenly
 think that, considering an oracle
evaluating propositional formulas in polynomial time, we could  immediately  define a ParTM solving SATISFIABILIY in polynomial time
(by defining, for  instance, instructions to set values $0$ and $1$ to propositional variables,   invoking the oracle to simultaneously evaluate
all possible values of the formula, and then using instructions with inconsistent conditions to establish whether any value $1$ was obtained).
However,  this is not the case, because only  parallelizable TMs can be taken as oracles.

As proven in Theorem \eqref{eq-comp-temp-partm-dtm}, ParTMs can be efficiently simulated
by DTMs. Then, if we had a ParTM solving SATISFIABILITY in polynomial time, this would lead to the surprising result that $P = NP$.
To avoid such a  mistake, we have to take into account the restriction
of parallelizability imposed to oracles in our model: If we had a parallelizable TM to
evaluate propositional formulas, it would  be easy to define a ParTM solving SATISFIABILIY in polynomial time and,
by Theorem \eqref{eq-comp-temp-partm-dtm}, we would  conclude $P = NP$. This only
shows the difficulty (or impossibility) in defining a parallelizable TM able to  evaluate propositional formulas.

On the other hand, Grover's quantum search algorithm and its  proven  optimality (see \cite[Chap. 6]{Nielsen-Chuang-2000}) implies
the non-existence of a `naive' search-based method to determine whether a function is constant or not in a time
less than $O(\sqrt{2^n})$.  This shows  that, in order to  take advantage of 
parallelism, inconsistency conditions on instructions featured by  ParTMs is a
more powerful property than quantum interference. However, in the case of ParTMs, this feature does not allow us to
define more efficient algorithms than otherwise defined by classical means. The reason is that the different paths of computations  may mix  in this model, and consequently 
we have to impose the parallelizability  restriction on oracles.
In the EParTMs model defined in the next section, different paths of computation do not mix indiscriminately as in ParTMs. Thus, no restrictions on the oracles are necessary, and, as it is shown in Theorem \eqref{EParTM-csat},
this new model of computation solves all NP-problems in polynomial time. This result shows that conditions of inconsistency on the instructions
are a really efficient method to take advantage of parallelism, and that this mechanism is more powerful than quantum interference.

\subsubsection{Simulating Entangled States and Relative Phases}\label{sim-ent-states-rel-phases}

In quantum theory, if we have $n$ physical systems with state
spaces $H_1, \ldots, H_n$ respectively, the system composed by
the $n$ systems has the state space $H_1 \otimes
\ldots \otimes H_n$ (in this case, $\otimes$ represent the
tensor product between state spaces) associated to it. Moreover, if we have
that the states of the $n$ component systems are respectively
$\ket{\psi_1}, \ldots, \ket{\psi_n}$, then the state of the
composed  system is $\ket{\psi_1} \otimes \ldots \otimes
\ket{\psi_n}$. However, there are states in  composed  systems that
cannot be described as tensor products of the states of the
component systems; these states are known as \emph{entangled
states}. An example of a two qubit entangled state is $\ket{\psi}
= \frac{1}{\sqrt{2}} (\ket{00} + \ket{11})$.

Entangled states enjoy the property that a measurement of one state in the
component system affects the state of other component systems,
even when systems are spatially separated. In this way, singular (one particle)
systems lose identity, because their states are  only describable in
conjunction with other systems. Entanglement is one of the
more (if not the most) puzzling characteristics of quantum
mechanics, with no  analogue in classical physics. Many quantum
computing researchers think that entangled states play a
definite role in the definition of efficient quantum algorithms,
but this is not a completely established fact; any elucidation
about this would be of great relevance. In this direction, we are going
to show how the concept of entanglement can be expressed in
logical terms, and we will define a new model of paraconsistent
TMs (EParTMs) in which uniform entangled states are well
represented.

As mentioned before, choice functions over the different
elements (state, position and symbol on the  cells of the tape) of
a ParTM, in a given instant of time $t$, determine a  classical TM
configuration. Then, a configuration of a ParTM can be viewed as
a uniform superposition of classical TM configurations where all
combinations of singular elements are taken into account.
Ignoring amplitudes, the tensor product of composed systems coincides with
all combinations of the basis states present (with amplitude greater than $0$) in the component systems.
For instance, if a system $S_1$ is in state $\ket{\psi_1} =
\ket{a_{i_1}} + \ldots + \ket{a_{i_n}}$ and a system $S_2$ is in
state $\ket{\psi_2} = \ket{b_{j_1}} + \ldots + \ket{b_{j_m}}$,
then the composed system  of $S_1$ and $S_2$ is in state
$\ket{\psi_{1,2}}  = \ket{a_{i_1} b_{j_1}} + \ldots +
\ket{a_{i_1} b_{j_m}} + \ldots + \ket{a_{i_n} b_{j_1}} + \ldots +
\ket{a_{i_n} b_{j_m}}$. This rule can be applied $n - 1$ times to
obtain  the state of a system composed by  $n$ subsystems.
Consequently, just by interpreting multiplicity of elements as superposed
states, ParTMs cannot represent entangled states, because all of
their configurations can be expressed as tensor products of
their singular elements. This is why we define the new model
of EParTMs, or ``entangled paraconsistent TMs'' (cf.
Definition~\ref{EParTM}).

In a ParTM configuration all combinations of its singular elements
are taken into account in the execution of its instructions (and also in the reading of the results).
This is because the logic $LFI1^*$, used in
the definition of the model,  validates the  \emph{rule  of
separation}  (i.e.  $\vdash_{\text{LFI1}^*} A \wedge B$ implies
$\vdash_{\text{LFI1}^*} A$ and $\vdash_{\text{LFI1}^*} B$) and the
\emph{rule of adjunction} (i.e.  $\vdash_{\text{LFI1}^*} A$ and
$\vdash_{\text{LFI1}^*} B$ implies $\vdash_{\text{LFI1}^*} A
\wedge B$). Then, for instance, if
$\Delta^{\star}_{\text{LFI1}^*}(\mc{M}(n))\vdash Q_1(\overline{t},
\overline{x}) \wedge S_1(\overline{t}, \overline{x})$ and
$\Delta^{\star}_{\text{LFI1}^*}(\mc{M}(n))\vdash Q_2(\overline{t},
\overline{x}) \wedge S_2(\overline{t}, \overline{x})$, it is also
possible to deduce $\Delta^{\star}_{\text{LFI1}^*}(\mc{M}(n))\vdash
Q_1(\overline{t}, \overline{x}) \wedge S_2(\overline{t},
\overline{x})$ and $\Delta^{\star}_{\text{LFI1}^*}(\mc{M}(n))\vdash
Q_2(\overline{t}, \overline{x}) \wedge S_1(\overline{t},
\overline{x})$.

By the previous explanation, if we want to define a model of
paraconsistent TMs where configurations are not totally mixed,
we have to consider a paraconsistent logic where the rule of
separation or the rule of adjunction are not both valid.
There exist non-adjunctive paraconsistent logics,\footnote{The
most famous of them is the \emph{discussive} (or
\emph{discursive}) logic $D2$, introduced by Stanis\l aw
Ja\'{s}kowski in \cite{Jaskowski-1948} and \cite{Jaskowski-1949},
with extensions to first order logic and with possible
applications in the axiomatization of quantum theory, cf.
\cite{daCosta-Doria-1995}.} but paraconsistent systems where the rule
of separation fails have never been proposed.
Moreover, despite the fact that non-adjunctive paraconsistent logics appear
to be an acceptable solution to avoid the phenomenon of complete mixing in ParTMs, the
notion of entanglement seems to be  more  related with the failure of
the rule of separation: Indeed, an entangled state describes the
`conjunctive' state of a composed system, but not the state of
each single subsystem. Thus, in order to define a model of
paraconsistent TMs better approaching the behavior of QTMs, we
first define a paraconsistent logic with a \emph{non-separable}
conjuction.

By following the ideas in \cite{Beziau-2002} (see also
\cite{Beziau-2005}), a paraconsistent negation $\neg_{\diamond}$ is defined into the
well-known modal system $S5$ (departing from classical negation $\neg$)
by $\neg_{\diamond} A \eqdef \diamond \neg A$
(some properties of this negation are presented in the referred
papers). We now  define a \emph{non-separable} conjunction $\wedge_{\diamond}$
into $S5$  by $A \wedge_{\diamond} B \eqdef \diamond (A \wedge B)$, where $\wedge$
is the classical conjunction. Some properties of this conjunction are
the following:
\begin{align}
    &\vdash_{S5} A \wedge_{\diamond} B \mbox{ does not imply neither }  \vdash_{S5} A \mbox{ nor } \vdash_{S5} B, \tag{$\wedge_{\diamond}1$}\label{ns-conj-prop-ns}\\
    &\vdash_{S5} A \mbox{ and } \vdash_{S5} B \mbox{ implies } \vdash_{S5} A \wedge_{\diamond}B, \tag{$\wedge_{\diamond}2$}\label{ns-conj-prop-ad}\\
    &\nvdash_{S5} \left(A \wedge_{\diamond}(B \wedge_{\diamond}C)\right) \leftrightarrow \left((A \wedge_{\diamond}B) \wedge_{\diamond}C\right), \tag{$\wedge_{\diamond}3$}\label{ns-conj-prop-nass}\\
    &\vdash_{S5} (A \wedge_{\diamond}B) \leftrightarrow (B \wedge_{\diamond}A), \tag{$\wedge_{\diamond}4$}\label{ns-conj-prop-conm}\\
    &\nvdash_{S5} \left((A \wedge_{\diamond}B) \wedge (C \wedge_{\diamond}D)\right) \rightarrow \left((A \wedge_{\diamond}D) \vee (C \wedge_{\diamond} B))\right)\tag{$\wedge_{\diamond}5$}\label{ns-conj-prop-comb}\\
    &\vdash_{S5} (A_1 \wedge_{\diamond}(A_2 \wedge \ldots \wedge A_n)) \leftrightarrow \diamond (A_1 \wedge \ldots \wedge A_n). \tag{$\wedge_{\diamond}6$}\label{ns-conj-prop-mult-conj}
\end{align}
Property \eqref{ns-conj-prop-ns} reflects  the non-separable
character of $\wedge_{\diamond}$,  while  \eqref{ns-conj-prop-ad} shows that
$\wedge_{\diamond}$ validates the rule of adjunction and
\eqref{ns-conj-prop-nass} grants the non-associativity
of $\wedge_{\diamond}$.  \eqref{ns-conj-prop-conm} shows that $\wedge_{\diamond}$ is
commutative, \eqref{ns-conj-prop-comb} is a consequence of
\eqref{ns-conj-prop-ns} related with the expression of entangled
states, and \eqref{ns-conj-prop-mult-conj} is a simple
application of the definition of $\wedge_{\diamond}$ which will be useful
below.

A paraconsistent non-separable logic, which we will call $PNS5$,  can be   `extracted'
from the modal logic $S5$ (as much as done for negation in  \cite{Beziau-2002})
 by inductively defining a translation  $\fp{*} ForPNS5 \to ForS5$
as:\footnote{Where $ForPNS5$ is the set of propositional formulas
generated over the signature $\sigma = \{\neg_{\diamond}, \wedge_{\diamond}, \vee,
\to\}$ (defined in the usual way) and $ForS5$ is the set of
formulas of $S5$.}
\begin{align*}
    &A^* = A \mbox{ if $A$ is atomic},\\
    &(\neg_{\diamond} A)^* = \diamond \neg (A)^*, \\
    &(A \wedge_{\diamond}B)^* = \diamond (A^* \wedge B^*),\\
    &(A \# B)^* = A^* \# B^* \mbox{ for $\# \in \{\vee, \to\}$};
\end{align*}
and by defining a  consequence relation in the wffs of  $PNS5$ as:
\begin{equation*}
    \Gamma \vdash_{PNS5} A \mbox{ iff } \Gamma^* \vdash_{S5} A^*,
\end{equation*}
where $\Gamma$ represents a subset of $ForPNS5$ and $\Gamma^* =
\{B^* | B \in \Gamma \}$. This translation completely specifies
$PNS5$ as   a sublogic of  $S5$  with the desired properties.

In the spirit of the LFIs (see
\cite{Carnielli-Coniglio-Marcos-2007}), we can define a
connective $\bullet$ of `inconsistency' in $PNS5$ by $\bullet A \eqdef A
\wedge_{\diamond}\neg_{\diamond} A$ (which is equivalent to $\diamond A
\wedge \diamond \neg A$ in $S5$), a connective $\circ$ of `consistency'
by $\circ A \eqdef \neg_{\diamond} \bullet A$ (which is equivalent
to $\square \neg A \vee \square A$ in $S5$), a classical negation $\neg$
by $\neg A \eqdef \neg_{\diamond} A \wedge_{\diamond}\circ A$ (which is
equivalent to $\diamond \neg A \wedge (\square \neg A \vee
\square A)$ in $S5$, entailing $\neg A$) and a
classical conjunction by $A \wedge B \eqdef (A \wedge_{\diamond} B) \wedge_{\diamond} (\circ A \wedge_{\diamond} \circ B)$
(which is equivalent to $\diamond (A \wedge B) \wedge (\square (A \wedge B) \vee
\square (A \wedge \neg B) \vee \square (\neg A \wedge B) \vee
\square (\neg A \wedge \neg B))$ in $S5$, entailing $A \wedge B$). Consequently, the ``explosion principles''
$(A \wedge \neg_{\diamond} A \wedge \circ A) \rightarrow B$, $(A \wedge_{\diamond} (\neg_{\diamond} A \wedge_{\diamond} \circ A)) \rightarrow B$,
$((A \wedge_{\diamond} \neg_{\diamond} A) \wedge_{\diamond} \circ A) \rightarrow B$ and
$((A \wedge_{\diamond} \circ_{\diamond} A) \wedge_{\diamond} \neg A) \rightarrow B$
are theorems of $PNS5$; in this way, $PNS5$ is a legitimate logic of formal inconsistency (cf. \cite{Carnielli-Coniglio-Marcos-2007}).
These definitions also allow us to fully embed classical
propositional logic into $PNS5$.

With the aim to use the logic $PNS5$ (instead of $LFI1^*$) in the definition of
EParTMs, we first need to extend $PNS5$ to first-order logic
with equality. This can be obtain by considering $S5Q^{=}$ (the first-order version of $S5$, with equality)
instead of $S5$ in the definition of the logic, and adjusting the translation function $*$ to deal with quantifiers and equality.
However, for the shake of simplicity, we will consider just $S5Q^{=}$ in the definition of the model,
and we will regard the connectives $\neg_{\diamond}, \wedge_{\diamond}, \bullet$
and $\circ$ as definitions into this logic. Then, we will substitute the underlying logic of intrinsic theories
$\Delta^{\star}_{FOL}(\mc{M}(\alpha))$ by  $S5Q^{=}$, and through the
Kripkean interpretation of $\Delta^{\star}_{S5Q^{=}}(\mc{M}(\alpha))$
theories, we will define what is a EParTM. Before that, we need
to identify  which kind of negation ($\neg$ or $\neg_{\diamond}$) and
conjunction ($\wedge$ or $\wedge_{\diamond}$) are adequate in each axiom
of $\Delta^{\star}_{S5Q^{=}}(\mc{M}(\alpha))$ (we will consider right-associative
conjunction, i.e., $A \wedge B \wedge C$ always mean $A \wedge (B
\wedge C)$;  this is necessary a proviso because of the non-associativity of
$\wedge_{\diamond}$, cf. property \eqref{ns-conj-prop-nass}):
\begin{enumerate}
    \item In axioms \eqref{existe-suc}-\eqref{antireflex-menorque}, negations and conjunctions are the classical ones;
    \item in axioms \eqref{ax-inst-i}-\eqref{ax-inst-iii}, the conjunction in the antecedent is $\wedge_{\diamond}$, (considering \eqref{ns-conj-prop-mult-conj}) only the first conjunction in the consequent is $\wedge_{\diamond}$ (other conjunctions are classical), and negation in \eqref{ax-inst-i} is classical;
    \item in axioms \eqref{init-conf} and \eqref{ax-t-0}, negations and conjunctions are the classical ones;
    \item in axiom \eqref{ax-t-halt}, only the conjunction in the antecedent is $\wedge_{\diamond}$, all other connectives are classical;
    \item in axioms \eqref{ax-unity-state} and \eqref{ax-unity-symbol}, all conjunctions are classical, but negations are $\neg_{\diamond}$ (except in $y \neq x$), and it is also necessary to add the connective $\diamond$ before the predicates $Q_i$ and $S_i$ into the antecedent of the axioms.
\end{enumerate}
We also need to define a notion of \emph{representation} for the
configurations of the TMs by worlds in a (possible-worlds) kripkean structure:
\begin{definition}
    Let $w$ be a world in a kripkean structure. If $Q_i(\overline{t}, \overline{x}), \ldots, S_{j_{-1}}(\overline{t}, -1), S_{j_{0}}(\overline{t}, 0), S_{j_{1}}(\overline{t}, 1), \ldots$ are valid predicates on $w$, we say that $w$ \emph{represents} a configuration for a TM $\mc{M}$ at time $\overline{t}$, and the configuration is given by the intended interpretation $I$ presented above.
\end{definition}

By considering the choices of connectives and the definition above,
worlds in the kripkean interpretation of
$\Delta^{\star}_{S5Q^{=}}(\mc{M}(\alpha))$ represent the parallel
computation of all possible computational paths of a NDTM
$\mc{M}$ for the input $\alpha$:
\begin{enumerate}
    \item By axiom \eqref{init-conf}, there will be a world $w_{0}$ representing the initial configuration of $\mc{M}(\alpha)$;
    \item by axioms \eqref{ax-inst-i}-\eqref{ax-inst-iii}, if $w_{t}$ represents a non-final configuration of $\mc{M}(\alpha)$ at time $t$, by any instruction $i_j$ to be executed at time $t$ (on such configuration), there will be a world $w_{t+1,j}$ representing a configuration of $\mc{M}(\alpha)$ at time $t+1$.
\end{enumerate}
Configurations represented by worlds for the same instant of time
$t$ can be considered \emph{superposed configurations}. In a
superposed configuration, a state on position $x$ and a symbol on
position $y$ are said to be \emph{entangled} if there exist $i, j, k, l$
($i \neq k$ and $j \neq l$) such that
$\Delta^{\star}_{S5Q^{=}}(\mc{M}(\alpha)) \vdash Q_i(\overline{t},
\overline{x}) \wedge_{\diamond}S_j(\overline{t}, \overline{y})$,
$\Delta^{\star}_{S5Q^{=}}(\mc{M}(\alpha)) \vdash Q_k(\overline{t},
\overline{x}) \wedge_{\diamond}S_l(\overline{t}, \overline{y})$ and
$\Delta^{\star}_{S5Q^{=}}(\mc{M}(\alpha)) \nvdash Q_i(\overline{t},
\overline{x}) \wedge_{\diamond}S_l(\overline{t}, \overline{y})$ or
$\Delta^{\star}_{S5Q^{=}}(\mc{M}(\alpha)) \nvdash Q_k(\overline{t},
\overline{x}) \wedge_{\diamond}S_j(\overline{t}, \overline{y})$. In a
similar way,  the notion of entangled symbols on positions $x$ and  $y$  can
also be defined.

In order to exemplify the definition above, consider the two qubits entangled state
$\ket{\psi} = \frac{1}{\sqrt{2}} (\ket{00} + \ket{11})$. Suppose the first qubit of $\ket{\psi}$
represents the state on position $x$ of an EParTM $\mc{M}$ (value $\ket{0}$ representing state $q_1$ and value $\ket{1}$ representing state $q_2$),
and the second qubit of $\ket{\psi}$ represents the symbol on position $y$ of $\mc{M}$
(value $\ket{0}$ representing symbol $s_1$ and value $\ket{1}$ representing symbol $s_2$).
Regarding only the state in position $x$ and the symbol in position $y$ of $\mc{M}$, state $\ket{\psi}$ represents a configuration of $\mc{M}$,
at time instant $t$, in which only the combinations $q_1 s_1$ and $q_2 s_2$ are possible, all other combinations being  impossible.
This is expressed in the theory $\Delta^{\star}_{S5Q^{=}}(\mc{M}(\alpha))$ by $\Delta^{\star}_{S5Q^{=}}(\mc{M}(\alpha)) \vdash Q_1(\overline{t},
\overline{x}) \wedge_{\diamond}S_1(\overline{t}, \overline{y})$,
$\Delta^{\star}_{S5Q^{=}}(\mc{M}(\alpha)) \vdash Q_2(\overline{t},
\overline{x}) \wedge_{\diamond}S_2(\overline{t}, \overline{y})$,
$\Delta^{\star}_{S5Q^{=}}(\mc{M}(\alpha)) \nvdash Q_1(\overline{t},
\overline{x}) \wedge_{\diamond}S_2(\overline{t}, \overline{y})$ and
$\Delta^{\star}_{S5Q^{=}}(\mc{M}(\alpha)) \nvdash Q_2(\overline{t},
\overline{x}) \wedge_{\diamond}S_1(\overline{t}, \overline{y})$.

Taking into account the definition of the inconsistency
connective in $S5Q^{=}$, as in the model of ParTMs, we can
define  conditions of inconsistency in the execution of
instructions in the EParTMs. In this case, by the definition of
the inconsistency connective in $S5Q^{=}$ and its  kripkean
interpretation, condition $q_i^\bullet$ will indicate that the
instruction will be executed only when at least two
configurations in the superposition differ in the current state
or position, while condition $s_j^\bullet$ will indicate that the
instruction will be executed only when at least two
configurations in the superposition differ in the symbol on the
position where the instruction can be executed.

A EParTM is then defined as:
\begin{definition}\label{EParTM}
A \emph{EParTM} is a NDTM such that:
\begin{itemize}
    \item When the machine reaches an ambiguous configuration with $n$ possible instructions to be executed, the machine configuration \emph{splits} into $n$ copies, executing a different instruction in each copy; the set of the distinct  configurations for an instant of time $t$ is called a \emph{superposed configuration};
    \item \emph{Inconsistency} conditions are allowed on the first two symbols of instructions (as indicated above);
    \item When there are no instructions to be executed (in any current configuration), the machine stops; at this stage the machine can be in a superposed configuration, each configuration in the superposition represents a result of the computation.
\end{itemize}
\end{definition}

Note that a EParTM parallelly performs all possible paths of
computation of a NDTM, and only such paths. This differs from   the
previous model of ParTMs, where the combination of actions of
different instructions had led to computational paths not possible
in the corresponding  NDTM.

Following \cite{Bennett-1973}, it is possible to define a
reversible EParTM for any EParTM without inconsistency
conditions in instructions.\footnote{In the case of EParTMs, it
is only necessary to avoid overlapping in the ranges of
instructions; the parallel execution of all possible instructions
in an ambiguous configuration does  not imply  irreversibility.}
This way, EParTMs almost coincide with  QTMs without amplitudes;
EParTMs  represent uniform superpositions with no
direct representation of the notion of relative phase,
but does allow conditions of inconsistency on the instructions.
As mentioned before, the notion of relative phase is a key
ingredient in allowing interference between different paths
of computation in QTMs, which is essential in order to take advantage of quantum
parallelism in the efficient solution of problems; however, this method has theoretical restrictions
which disable any definition of an efficient (polynomial time) quantum algorithm solving an NP-complete problem
by a naive search-based method (see \cite[Chap. 6]{Nielsen-Chuang-2000}). On the other
hand, conditions of inconsistency  on instructions provided by EParTMs
are  an efficient mechanism to accomplish actions
depending on different paths of computation. In fact, Theorem \ref{EParTM-csat} proves that
all NP-problems can be efficiently solved by EParTMs.

EParTMs  represent an abstract model of computation, independent of any  physical implementation.
However, if we think from the physical construction of EParTMs, quantum mechanics provides a way to implement the
simultaneous execution of different paths of computation, but does not provide
what it seems to be a simple operation over the superpositions obtained by quantum parallelism: The execution
of instructions depending on differences in elements on the superposed states (which correspond to
conditions of inconsistency on instructions of EParTMs). In this way, quantum mechanics does not supply a direct theoretical frame for the implementation
of EParTMs, but this definitely does not  forbid the possibility of a physical implementation of EParTMs (perhaps conditions of inconsistency could be
implemented by a sophisticated quantum physical procedure, or by a new physical theory).

We could also modify the definition of EParTMs to capture more  properties of QTMs. In this way,  conditions of inconsistency in instructions could be avoided,  and  a  notion of `relative phase'  could be introduced in EParTMs. This could  be achieved  by extending $S5Q^=$ with a new connective of possibility. Thus, the possibility connective of $S5$ (now denoted by $\diamond_1$) would  represent `positive configurations' and the other possibility connective ($\diamond_2$) would represent `negative configurations' (axioms establishing the behavior of $\diamond_2$ and their interrelation with other connectives would need to be added; in particular, combinations of connectives $\diamond_1$ and $\diamond_2$ would have to behave in an analogous way to combinations of symbols $+$ and $-$). The connective $\diamond_2$ could be used to define a new paraconsistent negation as well as   a new non-separable conjunction. Thus, by specifying which connectives would be used in each axiom, we could obtain a different definition of EParTMs. In this new definition, a concept of `interference' can be specified; equal configurations with the same possibility connective interfere constructively, while   equal  configurations with different possibility connectives interfere destructively. Although  details are  not given here,  this construction shows once more how we can define  computation models   with distinct  computational power by just substituting the  logic underlying theories $\Delta^{\star}_{FOL}(\mc{M}(\alpha))$. In this sense, computability can be seen as relative to logic. Alternatively, we can add a new element on the EParTMs: a sign indicating the `relative phase' of the configuration, and a new kind of instructions to change the relative phase.

\subsection{About the Power of ParTMs and EParTMs}\label{comp-power-partms-EParTMs}

In order to estimate the computational power of ParTMs and EParTMs,
we first define what the `deciding' of a language
(i.e. a set of strings of symbols $L \subset \Sigma^*$, where
$\Sigma$ is a set of symbols and $^*$ represents the Kleene
closure) means in these models of computation.
In the definition, we will consider multiple results in a computation as being possible
responses from which we have to randomly select only one.
We will also suppose that ParTMs and EParTMs have two distinguished
states: $q_y$ (the \emph{accepting state}) and $q_n$ (the
\emph{rejecting state}), and that all final states of the machine
(if it halts) are $q_y$ or $q_n$.
\begin{definition}
Let $\mc{M}$ be a ParTM (EParTM) and $x$ be a string of symbols in the input/output alphabet
of $\mc{M}$. We say that $\mc{M}$ \emph{accepts} $x$
with probability $\frac{m}{n}$ if $\mc{M}(x)$ halts in a
superposition of $n$ configurations and $m$ of them are in
state $q_y$; conversely, we say that $\mc{M}$ \emph{rejects} $x$
with probability $\frac{m}{n}$ if $\mc{M}(x)$ halts in a `superposition'
of $n$ configurations and $m$ of them are in state $q_n$.
Consequently, we say that $\mc{M}$ \emph{decides} a language $L$,
with error probability at most $1 - \frac{m}{n}$,
if for any string $x \in L$, $\mc{M}$ accepts $x$ with probability at least $\frac{m}{n}$,
and for any string $x \notin L$, $\mc{M}$ rejects $x$ with probability at least $\frac{m}{n}$.
\end{definition}

Bounded-error probabilistic time complexity classes are
defined for ParTMs and EParTMs as:
\begin{definition}
BParTM-PTIME (BEParTM-PTIME) is the class of
languages decided in \emph{polynomial time} by some ParTM (EParTM), with error probability at most $\frac{1}{3}$.
BParTM-EXPTIME (BEParTM-EXPTIME) is the class of
languages decided in \emph{exponential time} by some ParTM (EParTM), with error probability at most $\frac{1}{3}$.
\end{definition}
Space complexity classes can be defined in an analogous way, considering only the
largest space used for the different superposed configurations.

Now, we will prove that ParTMs are computationally equivalent to
DTMs, showing how to simulate the computation of ParTMs by DTMs
(Theorem \ref{eq-comp-partm-dtm}). As a consequence, we have
that the class of languages decided by both models of computation
is the same. It is obvious that computations performed by DTMs
can be computed also by ParTMs, because DTMs are particular cases
of ParTMs. What is surprising is that the simulation of ParTMs by
DTMs is performed with \emph{only} a polynomial slowdown in time
(Theorem \ref{eq-comp-temp-partm-dtm}) and a constant factor
overhead in space (direct consequence of the proof of Theorem
\ref{eq-comp-partm-dtm}). Theorems \ref{eq-comp-partm-dtm} and
\ref{eq-comp-temp-partm-dtm} are inspired in the simulation of
multi-tape TMs by one-tape TMs as presented in
\cite{Hopcroft-Motwani-Ullman-2001}, and show once more how
powerful the classical model of TMs is.

\begin{theorem}\label{eq-comp-partm-dtm}
    Any ParTM can be simulated by a DTM.
\begin{proof}
    Let $\mc{M}$ be a ParTM with $n$ states and $m$ input/output symbols. Define a DTM $\mc{M}'$ and suppose its tape is divided into $2n + m$ tracks. Symbols $1$ and $0$ on track $i$ ($1 \leq i \leq n$) and position $p$ of $\mc{M}'$ represent respectively that $q_i$ is or is not one of the states of $\mc{M}$ in position $p$. In a similar way, symbols $1$ and $0$ on track $j$ ($n + 1 \leq j \leq n + m$) and position $p$ of $\mc{M}'$ respectively represent the occurrence or non-occurrence of symbol $s_j$ on position $p$ of $\mc{M}$. Tracks $n + m + 1$ to $2n + m$ are used to calculate states resulting from the parallel execution of instructions in $\mc{M}$, and values on these tracks represent states of $\mc{M}$ in the same way as tracks $1$ to $n$. The symbol $\$$ is used on track $1$ of $\mc{M}'$ to delimitate the area where $\mc{M}$ is in any state (i.e., where any symbol $1$ appears on some track $i$ associated to states of $\mc{M}$). To simulate a step of the computation of $\mc{M}$, $\mc{M}'$ scans the tape between delimiters $\$$ in four times. In the first scan (from left to right), $\mc{M}'$ simulates the parallel execution of instructions where the action is a movement to right: In each position, $\mc{M}'$ writes values in tracks $n + m + 1$ to $2n + m$ in accordance with states `remembered' from the previous step and collects (in the state of the machine, depending on the content of tracks $1$ to $n + m$ and the instructions of movement to right of $\mc{M}$) the states to be written in the next position of the tape; $\mc{M}'$ also moves delimiters $\$$ if necessary. The second scan is similar to the first one, but in the opposite direction and simulating instructions of movement to the left, taking care in the writing  of values so as not to delete values $1$ written in the previous scan. In the third scan (from left to right), $\mc{M}'$ simulates the parallel execution of instructions where the action is the modification of symbols on the tape: In each position, depending on the content of tracks $1$ to $n + m$ and in accordance with the writing instructions of $\mc{M}$, $\mc{M}'$ writes values on tracks $n + 1$ to $n + m$ (corresponding to symbols written by instructions of $\mc{M}$) and also on tracks $n + m + 1$ to $2n + m$ (corresponding to changes of states from the writing instructions of $\mc{M}$, taking care in the writing  of values so as not to delete values $1$ written in the previous scans). Finally, $\mc{M}'$ performs a fourth scan (from right to left) copying values from tracks $n + m + 1$ to $2n + m$ on tracks $1$ to $n$ and writing $0$ on tracks $n + m + 1$ to $2n + m$.
\end{proof}
\end{theorem}

\begin{theorem}\label{eq-comp-temp-partm-dtm}
    The DTM of Theorem \ref{eq-comp-partm-dtm} simulates $n$ steps of the corresponding ParTM in time $O(n^2)$.
\begin{proof}
    Let $\mc{M}$ be a ParTM and $\mc{M}'$ be the DTM described in the proof of Theorem \ref{eq-comp-partm-dtm} such that  $\mc{M}'$   simulates the behavior of $\mc{M}$. After $n$ steps of computation, the leftmost state and the rightmost state of $\mc{M}$ cannot be separated by more than $2n$ cells, consequently this is the separation of $\$$ delimiters in the first track of $\mc{M}'$. In any scan of $\mc{M}'$, in the simulation of a step of computation of $\mc{M}$, $\mc{M}'$ has to move between $\$$ delimiters, and a writing operation can be performed in any position, thus any scan takes at most $4 n$ steps within the computation (ignoring steps due to scanning of delimiters $\$$ and  their possible relocation). Therefore, the simulation of the $n$ step in the computation of $\mc{M}$ takes at most $16 n$ steps, i.e., time $O(n)$. Consequently, for  the simulation of  $n$ steps in $\mc{M}$, $\mc{M}'$ requires no more than $n$ times this amount, i.e., time $O(n^2)$.
\end{proof}
\end{theorem}

\begin{corollary}
    The class of languages decided by ParTMs and DTMs are the same, and languages are decided with the same temporal and spatial complexity in both models.
\begin{proof}
    Direct consequence of theorems \ref{eq-comp-partm-dtm} and \ref{eq-comp-temp-partm-dtm}; it is only necessary to add another scan between delimiters $\$$ at the end of the simulation to search for an accepting state, finalizing $\mc{M}'$ in its accepting state if symbol $1$ is found in the track corresponding to the accepting state of $\mc{M}$, or finalizing $\mc{M}'$ in its rejecting state if no symbol $1$ is found in the track corresponding to the accepting state of $\mc{M}$. Clearly, this additional scan takes at most a polynomial number of steps (thus preserving the temporal complexity) and does not use new space (thus preserving the spatial complexity).
\end{proof}
\end{corollary}

For EParTMs the situation is different: The class
of languages decided in both models continues to be the same
(DTMs can simulate all paths of computation of a EParTM, writing
different configurations in separate portions of the tape and
considering the different configurations in the simulation of
instructions with inconsistency conditions), but all $NP$-problems
can be \emph{deterministically} (with error probability $0$) computed
in polynomial time by EParTMs (a direct consequence of Theorem
\ref{EParTM-csat}, since the satisfiability of propositional formulas
in conjunctive normal form  (CSAT) is $NP$-complete). Thus, time complexity
of EParTMs and DTMs  are equal only if $P = NP$, which is broadly
believed to be  false.

\begin{theorem}\label{EParTM-csat}
    CSAT is in BEParTMs-PTIME.
\begin{proof}
    It is not difficult to define a NDTM $\mc{M}$ deciding CSAT in polynomial time in which all computational paths have the same depth and finish in the same position of the tape. By considering $\mc{M}$ as a EParTM, all computational paths are performed in parallel, obtaining a superposition of configurations in which at least one of them is in state $q_y$ if the codified conjunctive normal form formula is satisfiable, or with all configurations in $q_n$ otherwise. Thus, by adding the instructions $i_{n+j}: q_y^\bullet s_j s_j q_y$ and $i_{n+m+j}: q_n^\bullet s_j s_j q_y$ to $\mc{M}$ (where $m$ is the number of input/output symbols of $\mc{M}$ and $1 \leq j \leq m$) we have the acceptance or rejection with probability 1.
\end{proof}
\end{theorem}

\section{Final Remarks}

In this paper, we generalize  a method for axiomatize Turing
machine computations not  only  with foundational aims, but
also  envisaging  new models of computation by  logical handling  (basically through the
substitution of the underlying logic  of the intrinsic theories in the computation),
showing a way in which logical representations can be used in the
construction of new concepts.

The  new  models of computation defined here use a  sophisticated
logical language  which permits us to express some  important features of
quantum computing. The  first model allows the simulation  of superposed
states by means of multiplicity of elements in TMs, enabling
the simulation of some quantum algorithms but unable to
speed up classical computation. In order to overcome
this weakness, we define a second model which is able to represent entangled
states, in this way, reaching an  exponential speed-up of an
$NP$-complete problem. Both models are grounded on paraconsistent
logic (LFIs). In particular, the  only element in the language
that cannot be  directly simulated in quantum computing is the
``inconsistency operator'' of  the second model. As this is a  key component
 in the efficiency of the whole model, an important
 problem  is to decide whether it can or cannot be characterized by
 quantum means.

In spite of \emph{paraconsistent computational theory} being only an
emerging field of research, we believe that this logic
relativization of the notion of computation is really  promising
in the search of efficient solutions to problems, particularly
helping in  the understanding of the role of quantum
features and indeterminism in computation processes.

\section*{Acknowledgements} 

This research was supported by FAPESP- Fundação de Amparo à Pesquisa do Estado
de São Paulo, Brazil, Thematic Research Project grant
2004/14107-2. The first author is also supported by a FAPESP
scholarship grant 05/04123-3, and the second by a CNPq (Brazil)
Research Grant 300702/2005-1.

\bibliographystyle{elsart-num}
%\bibliographystyle{plain}
%\bibliography{../juca}

\end{document}